\documentclass[a4paper,12pt]{article}

 \usepackage{amsmath}
 \usepackage{pstricks}
 \usepackage{pst-node}
 \usepackage[ansinew]{inputenc}
 \usepackage{amssymb,amsmath}
 \usepackage{amsfonts}
 \usepackage{epsfig}
\def\p{\partial}
\def\px{\partial_x}
\def\py{\partial_y}
\def\a{\alpha}
\def\b{\beta}
\def\g{\gamma}
\def\o{\omega}
\def\wt{\widetilde}

\author{Richard Beals$^{+}$ , Elena Kartashova$^*$ \\\\
$+$ Yale University, New Haven, CT,  USA \\
$*$ RISC, J.Kepler University, Linz, Austria\\
\\\\  e-mails:\\ richard.beals@yale.edu, lena@risc.uni-linz.ac.at
}

\title{Constructive factorization of LPDO in two variables}
\date{}
\begin{document}
 \maketitle

\abstract

We study conditions under which a partial differential operator of
arbitrary order $n$ in two variables or ordinary linear
differential operator admits a factorization with a first-order
factor on the left. The factorization process consists of solving,
recursively, systems of linear equations, subject to certain
differential compatibility conditions. In the generic case of
partial differential operators one does not have to solve a
differential equation.   In special degenerate cases, such as
ordinary differential, the problem is finally reduced to the
solution of some Riccati equation(s). The conditions of
factorization are given explicitly for second- and, and an outline
is given for the higher-order case.

 \section{Introduction}

Factorization of linear differential operators (LDO) is a very
well-studied problem and a lot of  pure existence theorems has
been proved (e.g. \cite{apel98}, \cite{olver93}, \cite{pom78}).
 The case of linear ordinary differential
operators (LODO) has been studied in more detail; various
algorithms are known for factoring LODOs over different
differential fields, e.g. \cite{bron94}, \cite{schw89},
\cite{tsarev96} and others. Probably the first algorithm for the
case of the simplest possible differential field (rational
functions) is described in \cite{beke94}. In this case
factorization into irreducible factors is known to be unique
\cite{loewy03}, \cite{loewy06} and in general, the problem finally
reduces to the
solution of a Riccati equation.\\

Much less is known about the factorization of linear partial
differential operators (LPDOs) which would appear to be much more
complicated than LODO. In several variables the naive definition
of factorization as representing a given $n$-th order operator
 as a composition of lower order operators does not have
 a nice uniqueness property;  this is illustrated by an
example of E. Landau  (formulated for left composition) \cite{landau}:
$$
(\px + 1)(\px + 1)(\px + x\py)=[\px^2 + x\px\py + \px +
(2+x)\py](\px + 1).
$$
 Nevertheless, some interesting
results have been obtained also in this direction;
recently Grigoriev and Schwarz \cite{grig04} have given an
algorithm for factoring an LPDO with separable symbol. Some
generalized definitions of factorization gave rise to another
factorization algorithms for factoring LPDO \cite{tsarev98} and
for factoring of systems of LPDEs with rational function
coefficients
\cite{tsarev01}, \cite{tsarev03}.\\

  It seems to be a general opinion
 that the naive factorization  -- even for second- and third order differential
 operators in two variables -- requires ``the solution of a partial
 Riccati equation, which in turn requires to solve
 a general first-order ODE and possibly an ordinary Riccati equation.
The bottleneck for designing
 a factorization algorithm for LPDO is the general first-order ODE which makes the full problem
 intractable at present because in general there are no solution algorithms
 available"~\cite{grig04}.\\

  We consider it possible to construct an explicit
  algorithm which could be used for absolute factorization of LPDO
  of arbitrary order $n$ in two variables into linear factors.
   The word "absolute''  means that we do not
  {\it fix} the coefficient field from the very beginning and that
  our only demand on the coefficients is that they be smooth, i.e.
  they belong to an appropriate differential field.
The procedure proposed here is to find a first order {\it left}
  factor (when possible) in contrast to the
  use of {\it right} factorization, which is common in the papers of
  last few decades.
  Of course the existence of a certain right factor of a LPDO is
  equivalent to the existence of a corresponding left factor of
  the {\it transpose\/} of that operator, so in principle nothing
  is lost by considering left factorization.  Moreover taking
  transposes is trivial algebraically, so there is also nothing
  lost from the point of view of algorithmic computation.\\

  The use of left factorization makes all the algebraic
  calculations needed much easier. It also yields explicit
  necessary and sufficient conditions on the coefficients
  of the operator for the existence of a factor. The simple form of
  these conditions allows us one to see the principal
  difference between factorization of LPDOs and LODOs - in the first
  case the  factorization problem can be solved by  pure algebraic
  methods in the "generic" case; the LODO is one example of the
  various degenerate cases in which it is necessary to solve a Riccati
  equation in order to factor in two variables.\\

  In Sec.~2 the general idea of factorization (in the generic case) by purely
  algebraic methods is presented and explicit conditions of factorization
  for an LPDO of order 2 are written out. It is also shown how the
  possibility of pure algebraic
  factorization for LPDO does not contradict the necessity to
  solve a Riccati equation in order to factor an LOPO.
 Similar results for LPDO and LODO of order three
   are demonstrated in Sec.~3, while in
  Sec.~4  the general procedure for
  factorization for LPDO of arbitrary order $n$ is presented.
  Some interesting examples are
 constructed  in Sec.~5 and Sec.~6 contains a brief concluding discussion.

 \section{LPDO of order 2 in two variables}

We consider an operator
\begin{equation}\label{A2}
A_2=\sum_{j+k\le2}a_{jk}\px^j\py^k
=a_{20}\px^2+a_{11}\px\py+a_{02}\py^2+a_{10}\px+a_{01}\py+a_{00}.
\end{equation}
with smooth coefficients and look for a factorization
$$
A_2=(p_1\px+p_2\py+p_3)(p_4\px+p_5\py+p_6).
$$

Let us write down the equations on $p_i$ explicitly, keeping in
mind the rule of {\it left} composition, i.e. that $ \px (\alpha
\py) = \px (\alpha) \py +
\alpha \partial_{xy}.$\\

Then in all cases

$$
 \begin{cases}
  a_{20} &= \ p_1p_4\\
  a_{11} &= \ p_2p_4+p_1p_5\\
  a_{02} &= \ p_2p_5\\
  a_{10} &= \ \mathcal{L}(p_4) + p_3p_4+p_1p_6\\
  a_{01} &= \ \mathcal{L}(p_5) + p_3p_5+p_2p_6\\
  a_{00} &= \ \mathcal{L}(p_6) + p_3p_6
  \end{cases}
  \eqno (\it{2SysP})
 $$\\
where we use the notation $\mathcal{L} = p_1 \px + p_2 \py $.\\

We begin with some preliminary remarks.  To some extent the factorization
problem is to be treated as a local problem; for example if some coefficient
does not vanish identically, we may wish to restrict to a region where it
has no zeros.  Of course insofar as the formulas obtained are analytic
functions of the data, if the data are also analytic then the formulas
may continue globally.
With this in mind, we note that if the operator is not (globally) of order one,
then we may, after a linear change of variables if necessary, assume that
$$
a_{20}\ne 0.
$$
Then necessarily $p_1\ne 0$, and we may assume without loss of generality
that
$$
p_1=1.
$$
Then the first three equations of
 {\it 2SysP}, describing the highest order terms are
equations in the variables $p_2, p_4, p_5$ and to find them we
have to find roots of a quadratic polynomial.  Having computed $p_2, p_4,
p_5$ one can plug them into two next equations of {\it 2SysP} and
get a {\it linear} system of equations in two variables $p_3,p_6$
which can easily be solved if $\o$ is a simple root.  Then the last
equation of {\it 2SysP} will give us the {{\it condition} of factorization.\\

 In fact, at {\bf the first step} from

$$
  \begin{cases}
  a_{20} = p_4 \\
  a_{11} = p_2p_4+p_5 \\
  a_{02} = p_2p_5
  \end{cases}
$$\\
it follows that
$$
\mathcal{P}_2(-p_2):=  a_{20}(- p_2)^2 +a_{11}(- p_2) +a_{02} = 0
$$
The choice of different roots of the characteristic polynomial
$\mathcal{P}_2$ gives us different (potential) factorizations of the
initial operator.  (As we shall see, there may be no such factorization,
or there may be one with one choice of root and not with another choice.)
Let $\o$ be a root of $\mathcal{P}_2$, and take
$$
p_2=-\o , \quad \mathcal{L} = \px - \o \py.
$$
This leads to a linear system for $p_4$, $p_5$ with $\o$ as parameter:
$$
  \begin{cases}
  p_4=a_{20},\\
  -\o p_4+p_5=a_{11}
  \end{cases}
$$
or, in matrix form,

$$
\begin{bmatrix} 1&0\cr -\o&1\end{bmatrix}\begin{bmatrix}p_4\cr p_5\end{bmatrix}
=\begin{bmatrix}a_{20}\cr a_{11}\end{bmatrix};\qquad
\begin{bmatrix}p_4\cr p_5\end{bmatrix}=
\begin{bmatrix}1&0\cr\o &1\end{bmatrix}
\begin{bmatrix}a_{20}\cr a_{11}\end{bmatrix}.
$$\\
Thus
 $$
  \begin{cases}
p_1=1\\
p_2=-\o\\
p_4=a_{20}\\
p_5=a_{20} \o +a_{11}
  \end{cases}
  \eqno (2{\it Pol})
 $$\\

 At {\bf the second step}, from

$$
  \begin{cases}
  a_{10} &= \ \mathcal{L}(p_4)+p_3p_4+p_1p_6\\
  a_{01} &= \ \mathcal{L}(p_5)+p_3p_5+p_2p_6
  \end{cases}\\
 $$
 and from (2{\it Pol}) we get
 $$
  \begin{cases}
  a_{10} &= \ \mathcal{L} a_{20} +p_3 a_{20} +p_6\\
  a_{01} &= \ \mathcal{L}(a_{11}+a_{20} \o)+p_3( a_{11} + a_{20}
  \o)-
  \o p_6.
  \end{cases}
  \eqno (2{\it Lin})
 $$\\
If the root $\o$ is simple,
i.e.~$\mathcal{P}_2'(\o)=2a_{20}\o+a_{11}\ne 0$, then these
equations have the unique solution

\begin{eqnarray}
  p_3 =  \frac{\o a_{10}+a_{01} -\o\mathcal{L}a_{20}- \mathcal{L}(a_{20} \o+a_{11})}
{2a_{20}\o+a_{11}}\nonumber\\
  p_6 =\frac{ (a_{20}\o+a_{11})(a_{10}-\mathcal{L})a_{20}-a_{20}(a_{01}
 -\mathcal{L}(a_{20}\o+a_{11})}{2a_{20}\o+a_{11}}.\nonumber
  \end{eqnarray}
At this point all coefficients $p_1, p_2, ..., p_6$ have been computed.\\

At {\bf the third step} from

$$
a_{00} =  \mathcal{L}(p_6)+p_3p_6
$$
the corresponding {\it condition of factorization} can be written out
explicitly:

\begin{eqnarray}
a_{00} = \mathcal{L} \left\{
 \frac{\o a_{10}+a_{01} - \mathcal{L}(2a_{20} \o+a_{11})}
{2a_{20}\o+a_{11}}\right\}+ \frac{\o a_{10}+a_{01} -
\mathcal{L}(2a_{20} \o+a_{11})}
{2a_{20}\o+a_{11}}\times\nonumber\\
\times\frac{ a_{20}(a_{01}-\mathcal{L})(a_{20}\o+a_{11})+
(a_{20}\o+a_{11})(a_{10}-\mathcal{L}a_{20})}{2a_{20}\o+a_{11}},\nonumber
 \end{eqnarray}\\
and we have just to check it. Thus factorization in two variables,
when possible, can (generally) be accomplished by purely algebraic
means.\\

{\bf Remark}.  In the preceding derivation we assumed that $\o$ is
a simple root so that $(a_{11}+2a_{20}\o)\neq 0$.
Suppose now that the condition fails identically in some region.  Then
by eliminating $p_6$ from (2\it Lin}) we obtain the necessary condition
$$
a_{10}\o+a_{01}-\mathcal{L} (a_{2 0}\o+a_{11}) -\o \mathcal{L}a_{2
0}=0. \eqno(2{\it Degen})
$$
If this condition is satisfied, then we may use the first equation of
(2{\it Lin}) to determine $p_6$ from $p_3$.  Then the last equation of
(2{\it SysP}) gives a Riccati equation for $p_3$.\\

Let us look explicitly at the case of an ordinary differential
operator of second order as a particular case of our initial LPDO.
Here $\o$ is a double root of the characteristic equation and the
necessary condtion (2{\it Degen}) is necessarily satisfied.  In
fact
 {\it 2SysP} takes the form
$$
 \begin{cases}

 a_{20} &= \ p_4\\
 a_{10} &= \ \mathcal{L}(p_4)  + p_3p_4 + p_6\\
 a_{00} &= \ \mathcal{L}(p_6) + p_3p_6\\
  \end{cases}
  \eqno (\it{2SysO})
 $$\\
\noindent  i.e.~we have only three non-trivial equations.
The first equation shows again
 that $p_4=a_{20}$,  $\mathcal{L}p_4=\dot{a}_{20}$ is known,
 and $p_6$ can written out explicitly as
$p_6= (a_{10} -\dot{a}_{20} - a_{20})p_3$. Substitution of this
expression for $p_4$ into the equation for $a_{00}$ gives us
immediately
  \begin{equation}
 a_{00} = \dot a_{10}-\ddot a_{20}- a_{20}\dot p_3-\dot a_{20}p_3+(a_{10}-
\dot a_{20}-a_{20}p_3)p_3
 \end{equation}\\
which is the Riccati equation
 \begin{equation}
\dot{\psi} + \psi^2+\frac{2\dot{a_{20}}-a_{10}}{a_{20}} \psi
+\frac{a_{00}+\ddot a_{20}-\dot a_{10}}{a_{20}}=0,
\quad {\rm for} \quad p_3=\psi.\nonumber
 \end{equation}\\
  In order to be able to find the general solution  explicitly
  one has to know at least one of its solutions.
  Of course the solution will depend on the explicit form
  of $a_{ij} = a_{ij}(x)$. \\

\section{ LPDO of order 3 }

Now we consider an operator
\begin{eqnarray}\label{A3}
A_3=\sum_{j+k\le3}a_{jk}\partial_x^j\partial_y^k =a_{30}\p_x^3 +
a_{21}\p_x^2 \py + a_{12}\px \py^2 +a_{03}\p y^3\\ \nonumber +
a_{20}\p_x^2+a_{11}\px\py+a_{02}\py^2+a_{10}\px+a_{01}\py+a_{00}.
\end{eqnarray}

with smooth coefficients and look for a factorization
$$
A_3=(p_1\px+p_2\py+p_3)(p_4 \p_x^2 +p_5 \px\py  + p_6 \py^2 + p_7
\px + p_8 \py + p_9).
$$

The conditions of factorization are described
 by the following system:
 $$
  \begin{cases}
  a_{30} &= \ p_1p_4\\
  a_{21} &= \ p_2p_4+p_1p_5\\
  a_{12} &= \ p_2p_5+p_1p_6\\
  a_{03} &= \ p_2p_6\\
  a_{20} &= \ \mathcal{L}(p_4)+p_3p_4+p_1p_7\\
  a_{11} &= \ \mathcal{L}(p_5)+p_3p_5+p_2p_7+p_1p_8\\
  a_{02} &= \ \mathcal{L}(p_6)+p_3p_6+p_2p_8\\
  a_{10} &= \ \mathcal{L}(p_7)+p_3p_7+p_1p_9\\
  a_{01} &= \ \mathcal{L}(p_8)+p_3p_8+p_2p_9\\
  a_{00} &= \ \mathcal{L}(p_9)+p_3p_9
   \end{cases}
  \eqno (\it{3SysP})
 $$
with $\mathcal{L} = p_1 \px + p_2 \py $.\\

Once again we may assume without loss of generality that the
coefficient of the term of highest order in $\p_x$ does not
vanish, and that the linear factor is normalized:
$$
a_{30}\ne 0,\qquad p_1=1.
$$
The first four equations of
 {\it 3SysVar} describing the highest order terms are
equations in the variables $p_2, p_4, p_5, p_6$.  Solving these
equations requires the choice of a root $-p_2$ of a certain polynomial of
third degree.  Once this choice has been made, the remaining
top order coefficients $p_4, p_5, p_6$ are easily found.
The top order coefficients can now be plugged into the next four
equations of {\it 3SysP}.  The first three of these four equations
will now be a {\it linear} system of equations in the variables
$p_3,p_7, p_8$ which is easily solved.
The next equation is now a {\it linear} equation
on variable $p_9$ which means that all variables $p_i, i=1,...,9$
have been found. The last two equations of {\it 3SysP} will give
us then the {\it conditions  of
factorization}.\\

 Namely, at {\bf the first step} from

$$
  \begin{cases}
  a_{30} = p_4 \\
  a_{21} = p_2p_4+p_5 \\
  a_{12} = p_2p_5+p_6\\
  a_{03} = p_2p_6
  \end{cases}\\
$$
it follows that
$$
\mathcal{P}_3(-p_2):=  a_{30}(-p_2)^3 +a_{21}(- p_2)^2 +
a_{12}(-p_2)+a_{03}=0.
$$
As for the case of second order, taking $p_2=-\o$,
where $\o$ is a
root of the characteristic polynomial $\mathcal{P}_3$
we get a linear system in
$p_4, p_5, p_6$ with $\o$ as parameter.  Then again
$$
p_2=-\o, \qquad
 \mathcal{L} = \px - \o \py,
$$
which leads to

$$
  \begin{cases}
  a_{30} = p_4 \\
  a_{21} = -\o p_4 + p_5 \\
  a_{12} = -\o p_5+p_6
  \end{cases}\\
$$
i.e.

$$
\begin{bmatrix} 1&0&0\cr -\o&1&0\cr 0&-\o&1\end{bmatrix}\begin{bmatrix}p_4\cr p_5\cr p_6\end{bmatrix}
=\begin{bmatrix}a_{30}\cr a_{21}\cr a_{12}\end{bmatrix};\qquad
\begin{bmatrix}p_4\cr p_5\cr p_6\end{bmatrix}=
\begin{bmatrix}1&0&0\cr\o &1&0 \cr \o^2& \o& 1 \end{bmatrix}
\begin{bmatrix}a_{30}\cr a_{21}\cr a_{12} \end{bmatrix}.
$$

Thus

 $$
  \begin{cases}
p_1=1\\
p_2=-\o\\
p_4=a_{30}\\
p_5=a_{30} \o+a_{21}\\
p_6=a_{30}\o^2+a_{21}\o+a_{12}.
  \end{cases}
  \eqno (3{\it Pol})
 $$

 At {\bf the second step}, from

$$
  \begin{cases}
  a_{20} &= \ \mathcal{L}(p_4)+p_3p_4+p_1p_7\\
  a_{11} &= \ \mathcal{L}(p_5)+p_3p_5+p_2p_7+p_1p_8\\
  a_{02} &= \ \mathcal{L}(p_6)+p_3p_6+p_2p_8\\
  \end{cases}\\
 $$
 and (2{\it Pol}) we get

 $$
  \begin{cases}
  a_{20}-\mathcal{L} a_{30} &=\ p_3 a_{30} +p_7\\
  a_{11}-\mathcal{L}(a_{30} \o+a_{21}) &=\ p_3(a_{30}\o+a_{21})- \o p_7+p_8\\
  a_{02}-\mathcal{L}(a_{30}\o^2+a_{21}\o+a_{12})&=\ p_3
(a_{30}\o^2+a_{21}\o+a_{12})-\o p_8.
  \end{cases}
  \eqno (3{\it Lin*})\\
 $$
As a linear system for $p_3$, $p_7$, $p_8$ this has determinant
$$
3a_{30}\o^2+2a_{21}\o+a_{12}=\mathcal{P}'(\o),
$$
so if $\o$ is a simple root the system has unique solution
\begin{eqnarray}
p_3 = \frac{\o^2 (a_{20} -\mathcal{L} a_{30})
+\o(a_{11}-\mathcal{L}(a_{30}\o+a_{21}+a_{02}))-\mathcal{L}
(a_{30}\o^2+a_{21}\o+a_{12})}{3a_{30}\o^2+2a_{21}\o+a_{12}};\nonumber\\
 p_7= \frac{a_{20}-\mathcal{L}a_{30}}{3a_{30}\o^2+2a_{21}\o+a_{12}}
-\frac{a_{30}}{3a_{30}\o^2+2a_{21}\o+a_{12}}\cdot p_3;\nonumber\\
 p_8= \frac{\o(a_{20}-\mathcal{L}a_{30})+a_{11}-\mathcal{L}(a_{30}\o+a_{21})}
{3a_{30}\o^2+2a_{21}\o+a_{12}}
-\frac{a_{30}\o +a_{21}}{3a_{30}\o^2+2a_{21}\o+a_{12}}\cdot p_3.\nonumber
\end{eqnarray}

 In order to find the last coefficient $p_9$ we use the next
equation of ({\it3SysVar}), namely:

$$
a_{10}  =  \mathcal{L}(p_7)+p_3p_7+p_1p_9, \eqno (3{\it Lin**}).
$$

At this point all coefficients $ p_i, i=1,...9$ have been computed,
under the assumption that $\o$ is a simple root. \\

At {\bf the third step} from\\

$$
\begin{cases}
a_{01} = \ \mathcal{L}(p_8)+p_3p_8+p_2p_9\\
a_{00} = \ \mathcal{L}(p_9)+p_3p_9
\end{cases}
$$
all the {\it necessary conditions} for factorization can be written
out.  We do not do so here because the formulas are
tedious and do not add anything to understanding the main
idea.   If the conditions are satisfied,  the explicit
factorization formulae could be written out as
for the second-order operator. The difference
is that in this case the polynomial defined by the
highest order terms is of degree 3
and we have not one but two conditions of factorization.\\

Regarding now  LODO as a particular case of LPDO let rewrite {\it
3SysP} with  $p_1=1$ as

 $$
  \begin{cases}
  a_{30} &= \ p_4\\
  a_{20} &= \ p_3p_4+p_1p_7+\dot{p_4}\\
  a_{10} &= \ p_3p_7+p_9+ \dot{p_7}\\
  a_{00} &= \ p_3p_9+ \dot{p_9}.
   \end{cases}
     \eqno (\it{3SysO})
 $$ \\
As in the second order case $\o=0$ is a multiple root
of $\mathcal{P}_3$.
After eliminating  $p_4=a_{30}$ and solving the second
and third equations for $p_7$ and $p_9$, respectively,
in terms of $p_3$,  we see finally that the
factorization problem for a third-order LODO is
equivalent to the following system of first order nonlinear ODEs
for the single unknown function $p_3$:

 $$
  \begin{cases}
  a_{10} =   p_7  \big(a_{20}  - \dot{a}_{30} -  p_7\big)/a_{30}+p_9+\dot{p_7}\\
  a_{00} = p_9 \big(a_{20}  - \dot{a}_{30} -  p_7\big)/a_{30} +\dot{p_9}.
   \end{cases}
 $$ \\
or, equivalently, a certain second-order nonlinear ODE.\\

 \section{LPDO of arbitrary order}

 Now consider the factorization problem for the LPDO of order $n$
 in the general form
\begin{equation}\label{An}
A_n=\sum_{j+k\le n}a_{jk}\partial_x^j\partial_y^k =(p_1\partial_x
+ p_2 \partial_y + p_3)\Big(\sum_{j+k <
n}p_{jk}\partial_x^k\partial_y^j\Big)
\end{equation}
and use the same assumptions and notations as above:
\begin{itemize}
\item{}some $a_{jk}$ with $j+k=n$ is not $=0$, so \item{} after a
linear change of variables one can assume that $a_{n0} \neq 0$,
\item{} set $p_1 =1$ and $p_{jk}=0$ if $j<0$ or $k<0$,
\item{}denote $p_2$ as $-\o$ and
$\mathcal{L}=p_1\px+p_2\py=\px-\o\py$.
\end{itemize}
Then the equations describing terms of highest order take the form
$p_1p_{j-1,k}+p_2p_{j,k-1} = a_{jk}$, i.e.
\begin{equation}\label{High}
p_{j-1,k}-\o p_{j,k-1}=a_{jk}, \quad j+k=n,
\end{equation}
while the equations of lower order take the form
\begin{equation}\label{Low}
\mathcal{L}p_{jk}+ p_3p_{jk}+ p_{j-1,k}-\o p_{j,k-1}=a_{jk}, \quad
j+k<n
\end{equation}
Keeping in mind that $j=n-k$ for the first $n$ equations \eqref{High}
let us rewrite them as
$$p_{n-1-k,k}-\o p_{n-k, k-1}= a_{n-k,k}, \quad k=0,1,...,n$$
 or in matrix form as
\begin{equation}\label{Highmatrix}
\begin{bmatrix} 1&0&0&...&0\cr -\o&1&0&...&0\cr 0&-\o&1&...&0\cr ...\cr 0&0&...& -\o&1\end{bmatrix}
\begin{bmatrix}p_{n-1,0}\cr p_{n-2,1}\cr p_{n-3,2}\cr ... \cr p_{0,n-1}\end{bmatrix}=
\begin{bmatrix}a_{n,0}\cr a_{n-1,1}\cr a_{n-2,2}\cr ... \cr a_{1,n-1}\end{bmatrix}.
\end{equation}

If we knew $\o$, the unique solution could be written out as
$$
\begin{bmatrix}p_{n-1,0}\cr p_{n-2,1}\cr p_{n-3,2}\cr ... \cr p_{0,n-1}\end{bmatrix}=
\begin{bmatrix} 1&0&0&...&0\cr \o&1&0&...&0\cr \o^2&\o&1&...&0\cr ...\cr \o^{n-1}&
\o^{n-2}&...& \o&1\end{bmatrix}
\begin{bmatrix}a_{n,0}\cr a_{n-1,1}\cr a_{n-2,2}\cr ... \cr
a_{1,n-1}\end{bmatrix},
$$
so
\begin{equation}\label{High2}
p_{n-k,k}=a_{n,0}\o^k+a_{n-1,1}\o^{k-1}+\dots+a_{n-k,k}.
\end{equation}
The equation with $k=n$, together with
$\o p_{0,n-1}=a_{0,n}$ gives us finally
$$
\mathcal{P}_n(\o):=a_{n,0}\o^{n}+a_{n-1,1}\o^{n-1}+...+a_{1,n-1}\o
+ a_{0,n}=0.
$$
Choosing $\o$ to be any root of this polynomial, we get then a
unique
determination of $p_{j,k}$ for $j+k=n$.\\

Consider now the $n$ equations of \eqref{Low} corresponding to the
terms of order $n-1$, i.e. the case $j+k=n-1$:
\begin{equation}\label{Low1}
p_{n-2-k,k}-\o p_{n-k-1,k-1}= b_{n-1-k,k}-p_3 p_{n-1-k,k},
\end{equation}
where the functions $p_{n-1-k,k}$ and
$$
b_{n-1-k,k}= a_{n-1-k,k}-\mathcal{L}p_{n-1-k,k}
$$
are known already.  In matrix form these equations are
$$
\begin{bmatrix}p_{n-1,0}&1&0&0&...&0\cr
p_{n-2,1}&-\o&1&0&...&0\cr
p_{n-3,2}&0&-\o&1&...&0\cr ...\cr
p_{0,n-1}&0&0&...& 0&-\o\end{bmatrix}
\begin{bmatrix}p_3\cr p_{n-2,0}\cr p_{n-3,1}\cr ...
\cr p_{0,n-2}\end{bmatrix}=
\begin{bmatrix}b_{n-1,0}\cr b_{n-2,1}\cr b_{n-3,2}
\cr ... \cr b_{0,n-1}\end{bmatrix}
$$
Inductively, the determinant of a matrix of this form is
$$
(-1)^{n-1}[p_{n-1,0}\o^{n-1}+p_{n-2,1}\o^{n-2}+\dots +p_{0,n-1}].
$$
Equations \eqref{High2} imply that the term in brackets is
$$
na_{n,0}\o^{n-1}+(n-1)a_{n-2,1}\o^{n-2}+\dots+a_{1,n-1}=\mathcal{P}'_n(\o).
$$
Therefore if $\o$ is a simple root, the equations
\eqref{Low1} for $p_3$ and the $p_{n-2-k,k}$ have a unique solution.
Multiplying the $j$-th equation of the system by $\o^{n-j}$ and
adding, we determine
\begin{equation}\label{p3}
\mathcal{P}_n'(\o)\,p_3=b_{n-1,0}\o^{n-1}+b_{n-2,1}\o^{n-2}+
\dots + b_{0,n-1}.
\end{equation}
If $\o$ is a simple root, then this equation determines $p_3$,
and we may determine the $p_{ni2-k,k}$ by rewriting the first
$n-1$ of these equations with the $p_3$ terms on the right hand
side, so that the solution for the $p_{n-2-k,k}$ in terms of
$p_3$ and other known functions is
\begin{equation}\label{matrixLow}
\begin{bmatrix}p_{n-2,0}\cr p_{n-3,1}\cr p_{n-4,2}\cr ... \cr p_{0,n-2}\end{bmatrix}=
\begin{bmatrix} 1&0&0&...&0\cr \o&1&0&...&0\cr \o^2&\o&1&...&0\cr ...\cr \o^{n-2}&
\o^{n-3}&...& \o&1\end{bmatrix}
\begin{bmatrix}b_{n-1,0}-p_3 p_{n-1,0}\cr
b_{n-2,1}-p_3 p_{n-2,1}\cr b_{n-3,2}-p_3 p_{n-3,2}
\cr ... \cr b_{1,n-2}-p_3 p_{1,n-2}\end{bmatrix}
\end{equation}
as before.  \\

Continuing to assume that $\o$ is a simple root,
let us follow this process one more step.  The first $n-2$ of the
$n-1$ equations for the coefficients $p_{n-3-k,k}$ can be written
in a matrix form like \eqref{Highmatrix}, with known functions on
the right hand side, so the solution can be written down as above.
Thus at this point we have determined the coefficients $p_j$ of
the first factor and all coefficients $p_{jk}$, $j+k\le n-3$ of
the second factor.  However there is one more equation at this
step:
$$
-\o p_{0,n-3}=a_{1,n-3}-(\mathcal{L}+p_3)p_{1,n-3}.
$$\\
Since both sides have already been determined, this is a necessary
condition for the existence of a factorization with this choice of
root $\o$. Continuing, the same situation occurs at each step.
Thus if the condition is satisfied at one step we may proceed to
the next step, and either the process fails or there are exactly
$n-1$ polynomial conditions that are satisfied by the
coefficients, the (simple) root $\o$, and their derivatives.\\

Suppose we drop the assumption that $\o$ is a simple root and
assume that $\mathcal{P}'_n(\o)\equiv0$ in some region. Then the
equation \eqref{p3}, with zero right hand side,
 is the necessary and sufficient condition for solvability
of the full system of equations for $p_3$ and the $p_{n-2-k.k}$.
If this condition is satisfied, then we may make any choice of
$p_3$ and determine the $p_{n-2-k,k}$ from \eqref{matrixLow}. At
subsequent stages we will encounter factorization conditions that
are expressible as polynomial equations in $p_3$ and its
derivatives (with coefficients that are polynomials in the
$a_{jk}$ and their derivatives and in the root $\o$ and its
derivatives); these can be thought of as generalized Riccati
equations for $p_3$.\\

In summary: starting with a simple root of the polynomial
$\mathcal{P}_n(\o)$ in the generic case $a_{n,0}\ne 0$, the
factorization process proceeds algebraically (in the coefficients
$a_{jk}$, the root $\o$, and their derivatives) as far as it can,
as determined by $n-2$ polynomial obstructions.  Starting with a
double root, then subject to one such polynomial constraint, the
remaining
constraints are $n-2$ generalized Riccati equations for $p_3$.\\

We call this eliminating procedure {\it order-reduction} for
obvious reasons. In trying to factor an operator of order $n$
as a product of a first order operator and an operator of
order $n-1$, we look at the equations associated to the terms
of given degree of the operator of order $n$, in descending
order.  The first set of equations gives us the top order terms
of both factors (modulo choice of a root of the characteristic equation),
the next set determines the next lower order coefficients,
and so on.

\section{Some examples}

\subsection{General second-order LPDO}

\begin{itemize}

\item {\bf Example 5.1.1. Hyperbolic operator:}\\

       $a_{20}=0, a_{11}=1, a_{02} =0$, i.e.
       $ L_2 := \px \py + a_{10} \px + a_{01} \py +
       a_{00}.$\\

       In this case either of two different factorizations may be possible:\\
        \begin{itemize}
        \item {\bf (5.1.1.1 )} if $a_{00}= \px a_{10} + a_{10} a_{01}$,
       then
        $$L_2= \left(\px + a_{01} \right)\left( \py + a_{10}\right)$$

        or\\

        \item {\bf (5.1.1.2 )} if $a_{00}= \py a_{01} + a_{10} a_{01}$,
       then  $$L_2= \left(\py + a_{10} \right)\left(\px + a_{01}\right).$$

       \end{itemize}

        Consider the  case of such an operator known as a Poisson operator:
        $$ L_2 = \px \py+\frac{1}{x+y}(\alpha \px + \beta \py )+
         \frac{\gamma}{(x+y)^2}$$

         where $\a, \b, \g$  are constants.\\

        Due to {\bf (5.1.1.1)} and  {\bf (5.1.1.2)}
      factorization is only possible  if $\gamma= \alpha (\beta -
        1)$ or $\gamma= \beta (\alpha - 1)$ and its explicit forms
        are, respectively:

 $$L_2= \left(\px + \frac{\beta}{x+y}
       \right)\left(\py + \frac{\alpha}{x+y}\right)$$

or
 $$L_2= \left(\py + \frac{\alpha}{x+y}\right)
       \left(\px + \frac{\beta}{x+y}\right).$$

       We note that in the particular case  $\g=0$, the
       Poisson operator reduces to
       the operator considered in Example 4 from
\cite{grig04} and the results there can
       be obtained immediately
       from the formulas above.\\

\item {\bf Example 5.1.2. Parabolic operator}

The operator
 $$
 A_p=a_{20}\p_x^2+a_{11}\px\py+a_{02}\py^2+a_{10}\px+a_{01}\py+a_{00}
 $$
is parabolic only if $a_{11}^2-4a_{20}a_{02}=0$. This
condition means that the characteristic polynomial
$$
\mathcal{P}_2:=  a_{20} \o^2 + a_{11} \o + a_{02}
$$
has a double root $\o=-a_{11}/a_{20}$ and it follows from
the discussion in Sec.~2 that factorization requires solution
of a Riccati equation for $p_3$.\\

\item {\bf Example 5.1.3. Elliptic operator}

The operator
 $$
 A_e=a_{20}\p_x^2+a_{11}\px\py+a_{02}\py^2+a_{10}\px+a_{01}\py+a_{00}
 $$
(with real coefficients)
is elliptic by definition if $a_{11}^2-4a_{20}a_{02}<0$, so that
the roots of the characteristic polynomial $\mathcal{P}_2$ are
complex.  Therefore, the factorization is only possible
over a field that contains $\mathcal{C}$.

\end{itemize}

\subsection{Role of zero term; order of factors}

Consider the following second order operators in two variables:
\begin{equation}\label{Ac}
A_a=\px^2-\py^2+x\py+y\px+\tfrac14(y^2-x^2)+a,
\end{equation}
where $a$ is a real constant.

\begin{itemize}
\item {\bf Example 5.2.1}

 We look for a factorization in the
form
$$
A_a=(p_1\px+p_2\py+p_3)(p_4\px+p_5\py+p_6).
$$
Necessarily $p_1p_4=1$, and with no essential loss of generality
we take $p_1=p_4=1$.  Then it is not difficult to show that the
factorization can only have one of two forms:
\begin{equation}\label{plusminus}
A_a=L^+_aL^-_a=(\px+\py+b)(\px-\py+c)
\end{equation}
or
\begin{equation}\label{minusplus}
A_a=L^-_aL^+_a=(\px-\py+b)(\px+\py+c)
\end{equation}
It follows from the results above, and can be checked directly as
an exercise, that the following is true:\\

\begin{itemize}
\item The operator
$$
A_a=\px^2-\py^2+y\px +x\py+\tfrac14(y^2-x^2)+a,\quad
$$
does not have a factorization of either form \eqref{plusminus} or
\eqref{minusplus} unless $a=\pm 1$.

\item The operator
$$
A_1=\px^2-\py^2+x\py+y\px+\tfrac14(y^2-x^2)+1
$$
has a factorization of the form \eqref{plusminus} but has no
factorization of the form \eqref{minusplus}.

\item The operator
$$
A_{-1}=\px^2-\py^2+x\py+y\px+\tfrac14(y^2-x^2)-1
$$
has a factorization of the form \eqref{minusplus} but has no
factorization of the form \eqref{plusminus}.\\
\end{itemize}

In fact
\begin{equation}\label{A1}
A_1=\big[\px+\py+\tfrac12(y-x)\big]\,\big[\px-\py+\tfrac12(y+x)\big];
\end{equation}
and
$$
A_{-1}=\big[\px-\py+\tfrac12(y+x)\big]\,\big[\px+\py+\tfrac12(y-x)\big].
$$\\

This example shows that the existence of a factorization depends
crucially on the term of order zero, and also that an operator may
have a {\it left\/} factor of a certain form but no {\it right\/}
factor of the same form, or conversely.

As we have shown, in two variables the question of existence of a
first order {\it left\/} factor of a given form can be settled,
and the the factors calculated,  in a systematic way using
procedures that are simple and (in general) purely algebraic.\\

\item {\bf Example 5.2.2}

The question of existence of a first order {\it right\/} factor
can be reduced to the question for a left factor by taking the
formal transpose.  As an illustration, let us find a right factor
for operator $A_1$ in this way.  In order to take advantage of
what is already known, we make a coordinate change $y\to-y$, which
converts $A_1$ to
$$
\wt A_1 =\px^2-\py^2-y\px-x\py+\tfrac14(y^2-x^2)+1.
$$
The transpose is
$$
{\wt
A}_1^t=\px^2-\py^2+y\px+x\py+\tfrac14(y^2-x^2)+1=A_1=L_1^+(L_1^-).
$$
Therefore
\begin{equation}\label{tildeA}
\wt A_1=(L_1^-)^t(L_1^+)^t=(-L_1^-)^t(-L_1^+)^t.
\end{equation}
Now according to \eqref{A1},
$$
(-L_1^+)^t=-\big[\px+\py+\tfrac12(y-x)\big]^t=\px+\py-\tfrac12(y-x).
$$
Undoing the coordinate transformation $y\to-y$ converts this last
operator to
$$
\px-\py+\tfrac12(y+x).
$$
According to \eqref{tildeA}, this is a right factor for $A_1$,
which is consistent with \eqref{A1}.

\end{itemize}

\subsection{Classes of factorizable LPDOs}

Putting some restrictions on the $a_{i,j}$ as functions of $x$ and $y$,
one can describe {\it all} factorizable operators in a given class
of functions. Let us exhibit explicit factorizations for one case of
hyperbolic operators:
\begin{equation}
A_2 := \partial _{xx} - \partial_{y y} +
a_{10} \partial_{ x} + a_{01} \partial_{ y} +
       a_{00},
\end{equation}
 i.e. $a_{20}=1, a_{11}=0, a_{02}=-1.$

       In this case there are only  two possible factorizations:

\begin{itemize}
       \item{case 1}: if $\quad 4 a_{00} = 2 (\partial_x + \partial_y) (a_{10} +
       a_{01})+ (a_{10}^2 - a_{01}^2)$ holds, then we
       have the following factorization:
      \begin{equation}
 A_2  =   \left(\partial_x + \partial_y +\frac{a_{10} - a_{01}}{2}\right)
       \left(\partial_x - \partial_y +\frac{a_{10} +
       a_{01}}{2}\right)
       \end{equation}\\

       \item{case 2}: if $ \quad 4 a_{00} = 2 (\partial_x + \partial_y) (a_{10} -
       a_{01})+ (a_{10}^2 - a_{01}^2)$ holds, then we
       have the following factorization:
\begin{equation}
        A_2  =   \left(\partial_x + \partial_y +\frac{a_{10} + a_{01}}{2}\right)
       \left(\partial_x - \partial_y +\frac{ a_{10} -
       a_{01}}{2}\right).
       \end{equation}\\

\end{itemize}

Let us consider  as a simple example the case when the lower order
coefficients
$a_{10},a_{01},a_{00}$ are linear functions of one variable $x$,
i.e.
$$
 \begin{cases}
 a_{10} &= \ b_{10,1}x+b_{10,0},\\
 a_{01} &= \ b_{01,1}x+b_{01,0},\\
 a_{00} &= \ b_{00,1}x+b_{00,0}.
  \end{cases}
 $$\\
Then the condition of factorization (24) is

$$
4b_{00,1}x+4b_{00,0}=2b_{10,1}+2b_{01,1}+b_{10,1}^2x^2+2b_{10,1}b_{10,0}x+b_{10,0}^2-
b_{01,1}^2x^2 - 2b_{01,1}b_{01,0}x - b_{01,0}^2
$$
which leads to\\

zero term:

$$
4b_{00,0}=2b_{10,1}+2b_{01,1}+b_{10,0}^2- b_{01,0}^2,
$$

linear term:

$$
4b_{00,1}=2b_{10,1}b_{10,0} - 2b_{01,1}b_{01,0},
$$

quadratic term:

$$
0=b_{10,1}^2 - b_{01,1}^2.
$$\\
The last equality gives immediately $b_{10,1}=\pm b_{01,1}$. Let us
consider, for instance, the case $b_{10,1}= - b_{01,1}=b$.  Then

$$
\begin{cases}
 4b_{00,0}=b_{10,0}^2 -
b_{01,0}^2\\
2b_{00,1}=-b(b_{10,0}+b_{01,0}),\\

\end{cases}
$$
or in parametric form:

$$
\begin{cases}
b_{01,0}=\sqrt{2}t_1 \sinh{t_2}, \\
b_{10,0}=\sqrt{2}t_1 \cosh{t_2}, \\
b_{00,0}= \frac{1}{2}t_1^2,\\
b_{00,1}=-\sqrt{2}t_1 t_3(\sinh{t_2}+\cosh{t_2}),\\
b_{10,1}= t_3\\
b_{01,1}=-t_3\\
\end{cases}
$$
with arbitrary real parameters $t_1,t_2, t_3 \in R$.  Varying
these parameters one can get coefficients
$b_{01,0},b_{10,0},...,b_{00,0}$ and, therefore, coefficients of the
initial operator $a_{10},a_{01},a_{00}$, that  belong to some
prescribed field.  For instance, with $t_1, t_3 \in Q $ and $t_2=0$,

$$
\begin{cases}
b_{01,0}=0, \\
b_{10,0}=\sqrt{2}t_1, \\
b_{00,0}= \frac{1}{2}t_1^2,\\
b_{00,1}=-\sqrt{2}t_1 t_3,\\
b_{10,1}= t_3\\
b_{01,1}=-t_3\\
\end{cases}
$$
and we get 2-parameter factorization in $Q(\sqrt{2})$:

$$
 A_2  =   \left(\px + \py +\frac{2t_3 x + \sqrt{2}t_1}{2}\right)
       \left(\px - \py +\frac{\sqrt{2}t_1       }{2}\right)
$$\\
The case $b_{10,0}=
 b_{01,0}$ can be treated in a way analogous to the previous one and case 2
in a way analogous to case 1.
 Combining all the results one will give a description of all
 factorizable LPDO of the form (22) with corresponding
 restrictions on the coefficients.

 \section{Conclusions}

The elimination order-reduction procedure has been presented for
explicit factorization of LPDO of arbitrary order. It was shown
that in generic case, i.e. in case when characteristic polynomial
has a distinct root, this procedure is wholly algebraic and no
differential equation has to be solved. Otherwise the
factorization problem is equivalent to solving some Riccati
equation. It was also demonstrated that different types of LPDO
have different factorization properties, namely, algebraic
factorization of hyperbolic operators, when possible, could be
done over $\mathcal{R}$, while factorization of elliptic operators
is only possible over $\mathcal{C}$. In case of parabolic operator
no algebraic factorization is possible.

 \section{Acknowledgements}
 The authors are very much obliged to the organizing committee of the
 Workshop  ``Integrable Systems'' which provided an
excellent opportunity to meet and work
 together. We also highly appreciate the  hospitality of Centro
 Internacional de Ciencias, Cuernavaca, Mexico where most of this work was
 done. We express  special gratitude to
Prof.~A. B. Shabat and Prof.~S. P. Tsarev for very stimulating
discussions during the writing of this paper.
 Author$^*$ gratefully acknowledges support of the Austrian Science Foundation
(FWF) under projects SFB F013/F1304.

 \end{document}